\documentclass[graybox]{svmult}

\usepackage{mathptmx}       
\usepackage{helvet}         
\usepackage{courier}        
\usepackage{type1cm}        
\usepackage{makeidx}         
\usepackage{graphicx}        
\usepackage{multicol}        
\usepackage[bottom]{footmisc}

\makeindex             
\begin{document}

\title*{The evolution of the core mass function by gas accretion}
\author{Sami Dib}
\institute{Sami Dib \at Imperial College London, Blackett Laboratory, Prince Consort road SW7 2AZ, London, United Kingdom, \email{sdib@imperial.ac.uk}}
\maketitle

\abstract
{We show how the mass function of dense cores (CMF) which results from the gravoturbulent fragmentation of a molecular cloud evolves in time under the effect of gas accretion. Accretion onto the cores leads to the formation of larger numbers of massive cores and to a flattening of the CMF. This effect should be visible in the CMF of star forming regions that are massive enough to contain high mass cores and when comparing the CMF of cores in and off dense filaments which have different environmental gas densities.}

\section{The initial core mass function: gravoturbulent fragmentation} \label{sec1}
It is well established that star formation occurs in dense, gravitationally bound cores which are embedded in a complex structure of intersecting filaments within molecular clouds. Considerable efforts have been made over the last years in order to understand the origin of the initial mass distribution of dense cores (ICMF) (Padoan \& Nordlund 2002,PN02; Hennebelle \& Chabrier 2008). However, these ideas neglected the time evolution of the ICMF. PN02 showed that the ICMF can be described by  Eq.~\ref{eq1}:

\begin{figure}
\includegraphics{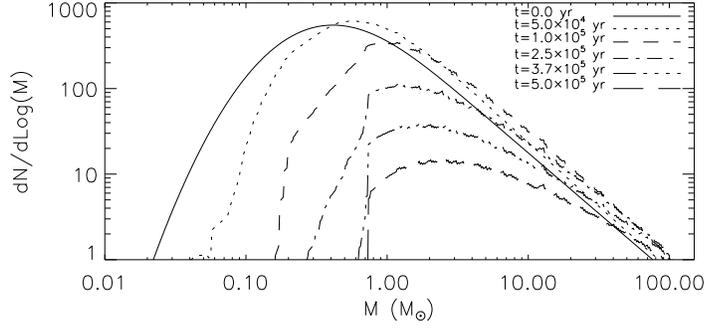}
\caption{Time evolution of the mass function of gravitationally bound cores in a molecular cloud. The CMF evolves over time as the result of gas accretion.}
\label{fig1}       
\end{figure}

\begin{equation}
N(M)~d{\rm log}M=f_{0}~\left[1+{\rm erf}\left( \frac{4 \ln M+\sigma_{d}^{2}}{2 \sqrt{2} \sigma_{d}}\right) \right]  M^{-3/(4-\beta)} d {\rm log}M,
\label{eq1}
\end{equation}    

where $\sigma_{d}$ is the width of the log-normal distribution of the density field and which is given by $\sigma_{d}={\rm ln}(1+\gamma^{2} {\cal M}^{2})$, where ${\cal M}$ is the Mach number, $\beta$ is the exponent of the kinetic energy power spectrum $E_{k} \propto k^{-\beta}$, and $f_{0}$ is a normalisation coefficient. 

\section{The evolution of the core mass function by accretion} \label{sec2}
Dib et al. (2010) argued that the ICMF will evolve in time as an unavoidable consequence of the cores growing in mass by gas accretion. They showed that the evolution of the CMF is described by the following time-dependent equation:

\begin{equation} 
\left(\frac{dN}{dt}\right) (M,t)=\nonumber \\
\left[-\left(\frac{\partial N}{\partial M} \right) \dot{M}-\left(\frac{\partial \dot{M}}{\partial M}\right) N\right] (M,t),
\label{eq2}
\end{equation}

where $\dot{M}$ is the accretion rate. In Dib et al. (2010), $\dot{M}$ had a time dependent component and the assumption was made that cores form uniformly over time with a prescribed core formation efficiency per unit time. They showed that a important consequence of Eq.~\ref{eq2} is the development of a larger fraction of massive cores and a flattening of the CMF (and IMF) at the high mass end. Here we show the evolution of the CMF for a single population of cores formed at $t=0$ and which accretes with the following accretion rate (Basu \& Jones 2004): $\dot{M}=\psi M^{2/3}$, where $\psi=(36 \pi)^{1/3} t_{d}^{-1}$ and $t_{d}=1/(G \rho_{ext})^{1/2}$ is the external medium crossing time and $\rho_{ext}$ is the external medium density. Fig.~\ref{fig1} displays the time evolution of the CMF (with $\gamma=0.5$, ${\cal M}=6$, $\rho_{ext}=8\times10^{-21}$ g cm$^{-3}$, $f_{0}=10^{2}$, $\beta=1.88$). The figure clearly shows the buildup of massive cores and the flattening of the CMF as time goes by.     

\begin{acknowledgement}
I acknowledge support from STFC grant ST/H00307X/1.  
\end{acknowledgement}

\begin{thebibliography}{99.}
\bibitem{science-journal} Dib, S., Shadmehri, M., Padoan, P., Maheswar, G., Ohja, D. K., Khajenabi, F.  MNRAS \textbf{405}, 401-- 420 (2010)
\bibitem{science-journal} Basu, S., Jones, C. E. MNRAS \textbf{347}, L47--L51 (2004) 
\end{thebibliography}
\end{document}